\documentclass[aps,preprintnumbers,eqsecnum,amsmath,amssymb,showpacs,nofootinbib]{revtex4}
\usepackage{graphicx}
\usepackage{dcolumn}
\usepackage{bm}
\usepackage{epsfig}

\begin{document} 
\title{\boldmath Narrow exotic tetraquark mesons in large-$N_c$ QCD}
\author{Wolfgang Lucha$^a$, Dmitri Melikhov$^{a,b,c}$, Hagop Sazdjian$^d$}
\affiliation{$^a$Institute for High Energy Physics,
Austrian Academy of Sciences, Nikolsdorfergasse 18, A-1050 Vienna,
Austria\\ $^b$D.~V.~Skobeltsyn Institute of Nuclear Physics,
M.~V.~Lomonosov Moscow State University, 119991, Moscow, Russia\\
$^c$Faculty of Physics, University of Vienna, Boltzmanngasse 5,
A-1090 Vienna, Austria\\ $^d$Institut de Physique Nucl\'eaire,
CNRS-IN2P3, Universit\'e Paris-Sud, Universit\'e Paris-Saclay,
91406 Orsay, France}
\date{\today}

\begin{abstract}
Discussing four-point Green functions of bilinear
quark currents in large-$N_c$ QCD, we formulate rigorous criteria
for selecting diagrams appropriate for the analysis of potential
tetraquark poles. We find that both flavor-exotic and cryptoexotic
(i.e., flavor-nonexotic) tetraquarks, if such poles exist, have a
width of order $O(1/N_c^2)$, so they are parametrically narrower
compared to the ordinary $\bar qq$ mesons, which have a width of
order $O(1/N_c)$. Moreover, for flavor-exotic states, the
consistency of the large-$N_c$ behavior of ``direct'' and
``recombination'' Green functions requires two narrow
flavor-exotic states, each coupling dominantly to one specific
meson-meson channel.
\end{abstract}

\maketitle

\section{Motivation}
To provide the theoretical understanding of exotic tetraquark
mesons, many candidates for which have been reported in the recent
years (see \cite{olsen,ali}), it is conventional to refer to QCD
with a large number of colors $N_c$ [i.e., ${\rm SU}(N_c)$ gauge
theory for large $N_c$] with a simultaneously decreasing coupling
$\alpha_s\sim1/N_c$ \cite{largeNc1,largeNc2}: at $N_c$-leading
order, large-$N_c$ QCD Green functions have only non-interacting
mesons as intermediate states; tetraquark bound states may emerge
only in $N_c$-subleading diagrams \cite{coleman}. For many years,
this fact was believed to provide the theoretical explanation of
the non-existence of exotic tetraquarks. However, as emphasized in
\cite{weinberg}, even if the exotic tetraquark bound states appear
only in subleading diagrams, the crucial question is the width of
these objects: if narrow, they might be well observed in nature.
The conclusion of \cite{weinberg} was that, if tetraquark states
exist, they may be as narrow as the ordinary mesons, i.e., have a
width $\sim1/N_c$. This issue has been further addressed in
\cite{knecht}, discussing the dependence of the width of
tetraquark mesons on their flavor structure. Finally,
\cite{maiani} reported for the cryptoexotic tetraquarks an even
smaller width of order $N_c^{-3}$.

Before drawing a conclusion about the width of a potential
tetraquark pole in large-$N_c$ QCD, it is mandatory to formulate
rigorous criteria for selecting those QCD diagrams that may lead
to the appearance of this pole: the crucial property of such
sequence of diagrams is the presence of four-quark intermediate
states and the corresponding cuts in the variable $s$ if one
expects to observe a tetraquark pole in $s$ \cite{cohen}. The
presence of such a four-particle $s$-cut should be established on
the basis of the Landau equations \cite{landau}. It turns out that
some of the diagrams attributed to the tetraquark pole in previous
analyses are lacking the necessary four-particle cut and thus may
not be related to the tetraquark properties. According to our
findings, the tetraquark width at large $N_c$ does not depend on
its flavor structure: both flavor-exotic and flavor-nonexotic
tetraquarks have the same width of order $1/N_c^2$.

We analyse four-point Green functions of bilinear quark currents
of various flavor content; any such function~depends on six
kinematical variables: the four momenta squared of the external
currents, $p_1^2$, $p_2^2$, $p_1^{\prime2}$, $p_2^{\prime2}$,
$p=p_1+p_2=p_1'+p_2'$, and the two Mandelstam variables $s=p^2$
and $t=(p_1-p_1')^2$. When selecting the diagrams which
potentially contribute to the tetraquark pole at $s=M_T^2$, we
apply the following two criteria:
\begin{enumerate}
\item 
The diagram should have a nontrivial (i.e., non-polynomial) dependence on the variable $s$.
\item
The diagram should have four-quark intermediate states and corresponding cuts starting at
$s=(m_1+m_2+m_3+m_4)^2$, where $m_i$ are the masses of the quarks
forming the tetraquark bound state. The presence or absence of
this cut is established by solving the Landau equations for the
corresponding diagram.
\end{enumerate}
Making use of the Landau equations is an unambiguous way to identify the set 
of QCD diagrams which have the four-quark cut; this is a necessary (although not a sufficient) 
condition that these diagrams contribute to the tetraquark pole. The four-quark cut should be 
present in the individual QCD diagrams that are appropriate for the tetraquark analysis, but of course 
this cut will be replaced in the complete Green functions by the tetraquark pole (in case it exists), 
plus the meson continuum, as soon as the infinite set of QCD diagrams is considered. 

Not all diagrams in the
perturbative expansion of Green functions satisfy the above
criteria, so we decompose these diagrams into two sets: diagrams
belonging to the first set either do not depend on $s$ or have no
four-quark cut in the $s$-channel and thus are not related to
tetraquarks; diagrams of the second set satisfy both of the two
above criteria~and thus contribute to the potential tetraquark
pole.

With the formulated criteria at hand, we discuss separately two
cases: tetraquarks of an exotic flavor content, i.e., built up of
quarks of four different flavors, $\bar q_1q_2\bar q_3q_4$, and
cryptoexotic tetraquarks, with flavor content $\bar q_1q_2\bar
q_2q_3$,~carrying the same flavor as ordinary mesons. The need for
separate treatment of flavor-exotic and cryptoexotic cases
arises~from the different topologies of the QCD diagrams emerging
for these two cases.

\section{Flavour-exotic tetraquarks}
Let us consider a bilinear quark current $j_{ij}=\bar q_i\hat O
q_j$ producing a meson $M_{ij}$ of flavor content $\bar q_iq_j$
from the vacuum, $\langle 0|j_{ij}|M_{ij}\rangle=f_{M_{ij}}$.
Here, $\hat O$ is a combination of Dirac matrices corresponding to
the meson's spin and parity. We~shall omit all Lorentz structures
as they are irrelevant for our analysis. At large $N_c$, the meson
decay constants $f_{M}$ scale~as $f_{M}\sim\sqrt{N_c}$.

In the case of four-point functions of bilinear currents involving
quarks of four different flavors, denoted by $1,2,3,4$, there are
two types of Green functions: the ``direct'' functions
$\Gamma^{\rm(dir)}_{{\rm I}}=\langle
j^\dagger_{12}j^\dagger_{34}j_{12}j_{34}\rangle$ and
$\Gamma^{\rm(dir)}_{{\rm II}}=\langle
j^\dagger_{14}j^\dagger_{32}j_{14}j_{32}\rangle$, and the
``recombination'' functions $\Gamma^{\rm(rec)}=\langle
j^\dagger_{12}j^\dagger_{34}j_{14}j_{32}\rangle$ and
$\Gamma^{\rm(rec)\dag}$.

\begin{figure}[!h]
\centering
\includegraphics[width=14cm]{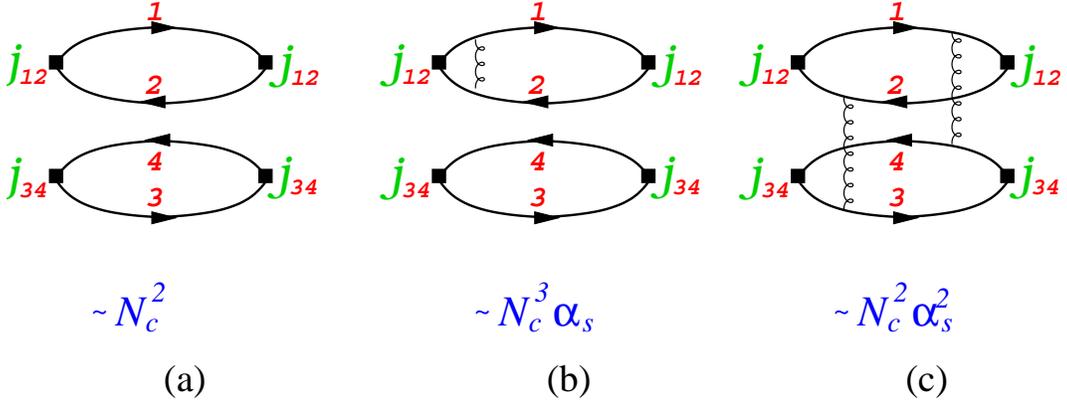}
\caption{\label{Fig:dir}Perturbative expansion of the Green functions $\Gamma^{\rm(dir)}_{I}$.}
\end{figure}

Figure~\ref{Fig:dir} shows the perturbative expansion of the
direct correlator $\Gamma^{\rm(dir)}_{{\rm I}}$. Similar diagrams
defined by evident flavor rearrangements describe the correlator
$\Gamma^{\rm(dir)}_{{\rm II}}$. Obviously, not all these diagrams
satisfy our above criteria for diagrams that potentially contain a
tetraquark pole. For instance, the diagrams in
Fig.~\ref{Fig:dir}(a,b) do not depend on $s$. The leading
large-$N_c$ diagram which depends on $s$ and also has a four-quark
$s$-cut is given by Fig.~\ref{Fig:dir}(c). The diagrams of this
type are therefore the leading large-$N_c$ diagrams of interest to us.

\begin{figure}[!t]
\centering
\includegraphics[width=14cm]{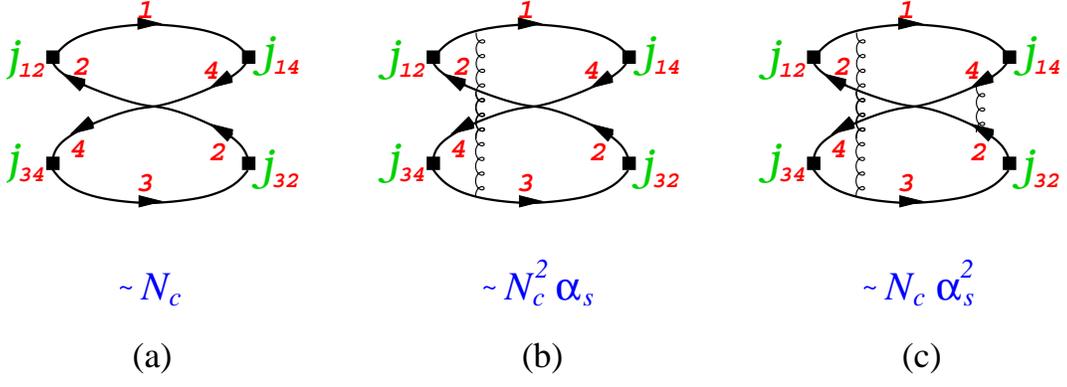}
\caption{\label{Fig:rec}Diagrams for the recombination Green function $\Gamma^{(\rm rec)}$.}
\end{figure}

The analysis of the recombination channel is a bit more involved:
among the diagrams in Fig.~\ref{Fig:rec}, the first two diagrams
(a,b) do depend on $s$; however, in spite of their appearance,
they have no four-quark cut. The easiest way to see this is to
redraw these diagrams as the usual box diagram (a) and the box
diagram with one-gluon exchange~(b). The $N_c$-leading diagram
exhibiting the four-quark cut is the nonplanar diagram in
Fig.~\ref{Fig:rec}(c). The four-quark cut with the threshold at
$s=(m_1+m_2+m_3+m_4)^2$ may be verified by solving the Landau
equations. We thus find the following $N_c$-leading behavior of
the Green functions $\Gamma_{T}$ potentially involving a pole
corresponding to some tetraquark $T$:
\begin{eqnarray}
\label{T1} \Gamma^{\rm(dir)}_{{\rm I},T}=\langle
j^\dagger_{12}j^\dagger_{34}j_{12}j_{34}\rangle=O(N_c^0),\quad
\Gamma^{\rm(dir)}_{{\rm II},T}=\langle
j^\dagger_{14}j^\dagger_{32}j_{14}j_{32}\rangle=O(N_c^0),\quad
\Gamma^{\rm(rec)}_T=\langle
j^\dagger_{12}j^\dagger_{34}j_{14}j_{32}\rangle=O(N_c^{-1}).
\end{eqnarray}
These diagrams potentially contain the tetraquark pole, although
the actual existence of this pole is still a conjecture. Now, let
us assume that narrow resonances (i.e., resonances with widths
vanishing for large $N_c$) show up at the~lowest possible $1/N_c$
order and that the resonance mass $M_T$ remains finite at large
$N_c$. The fact that direct and recombination amplitudes have
different behaviors in $N_c$ leads us to the conclusion that a
single pole is not sufficient and we need~at least two exotic
poles, denoted by $T_A$ and $T_B$: $T_A$ couples stronger to the
$M_{12}M_{34}$ channel, while $T_B$ couples stronger to the
$M_{14}M_{32}$ channel.

Truncating the poles corresponding to the external mesons and
retaining explicitly only the tetraquark poles, we~get
\begin{eqnarray}\label{T4b}\Gamma_{{\rm I},T}^{(\rm
dir)}&=&O(N_c^0)=f_M^4\left(\frac{|A(M_{12}M_{34}\to
T_A)|^2}{p^2-M^2_{T_A}}+\frac{|A(M_{12}M_{34}\to
T_B)|^2}{p^2-M^2_{T_B}}\right)+\cdots,\nonumber\\\Gamma_{{\rm
II},T}^{(\rm dir)}&=&O(N_c^0)=f_M^4\left(\frac{|A(M_{14}M_{32}\to
T_A)|^2}{p^2-M^2_{T_A}}+\frac{|A(M_{14}M_{32}\to
T_B)|^2}{p^2-M^2_{T_B}}\right)+\cdots,\\\Gamma_{T}^{(\rm
rec)}&=&O(N_c^{-1})=f_M^4\left(\frac{A(M_{12}M_{34}\to
T_A)A(T_A\to M_{14}M_{32})}{p^2-M^2_{T_A}}+\frac{A(M_{12}M_{34}\to
T_B)A(T_B\to M_{14}M_{32})}{p^2-M^2_{T_B}}\right)+\cdots.\nonumber
\end{eqnarray}
Taking into account that $f_M\sim\sqrt{N_c}$ and that we are
seeking tetraquarks with finite mass at large $N_c$, these
equations have the following solution:
\begin{align}
\label{T4b2}
A(T_A\to M_{12}M_{34})&=O(N_c^{-1}),\qquad&A(T_A\to
M_{14}M_{32})&=O(N_c^{-2}),\nonumber\\A(T_B\to
M_{12}M_{34})&=O(N_c^{-2}),\qquad&A(T_B\to
M_{14}M_{32})&=O(N_c^{-1}).
\end{align}
The widths $\Gamma(T_{A,B})$ of the states $T_A$ and $T_B$ are
determined by the dominant channel, which yields
$\Gamma(T_{A,B})=O(N_c^{-2})$.

So far we have ignored the mixing between $T_A$ and $T_B$.
Introducing their mixing parameter $g_{AB}$, we get additional
contributions to the above Green functions. Most restrictive for
$g_{AB}$ is the recombination function, for which mixing provides
the additional contribution
\begin{eqnarray}\label{mixing}
\Gamma_{T}^{(\rm rec)}=O(N_c^{-1})
=f_M^4\left(\frac{A(M_{12}M_{34}\to T_A)}{p^2-M^2_{T_A}}g_{AB}
\frac{A(T_B\to M_{14}M_{32})}{p^2-M^2_{T_B}}\right)+\cdots.
\end{eqnarray}
Equations (\ref{T4b2}) and (\ref{mixing}) restrict the behavior of
the mixing parameter to $g_{{}_{AB}}\le O(N_c^{-1})$. Thus, the
two flavor-exotic tetraquarks of the same flavor content do not
mix at large $N_c$.

\section{Cryptoexotic tetraquarks}
\begin{figure}[!t]\centering
\includegraphics[width=8cm]{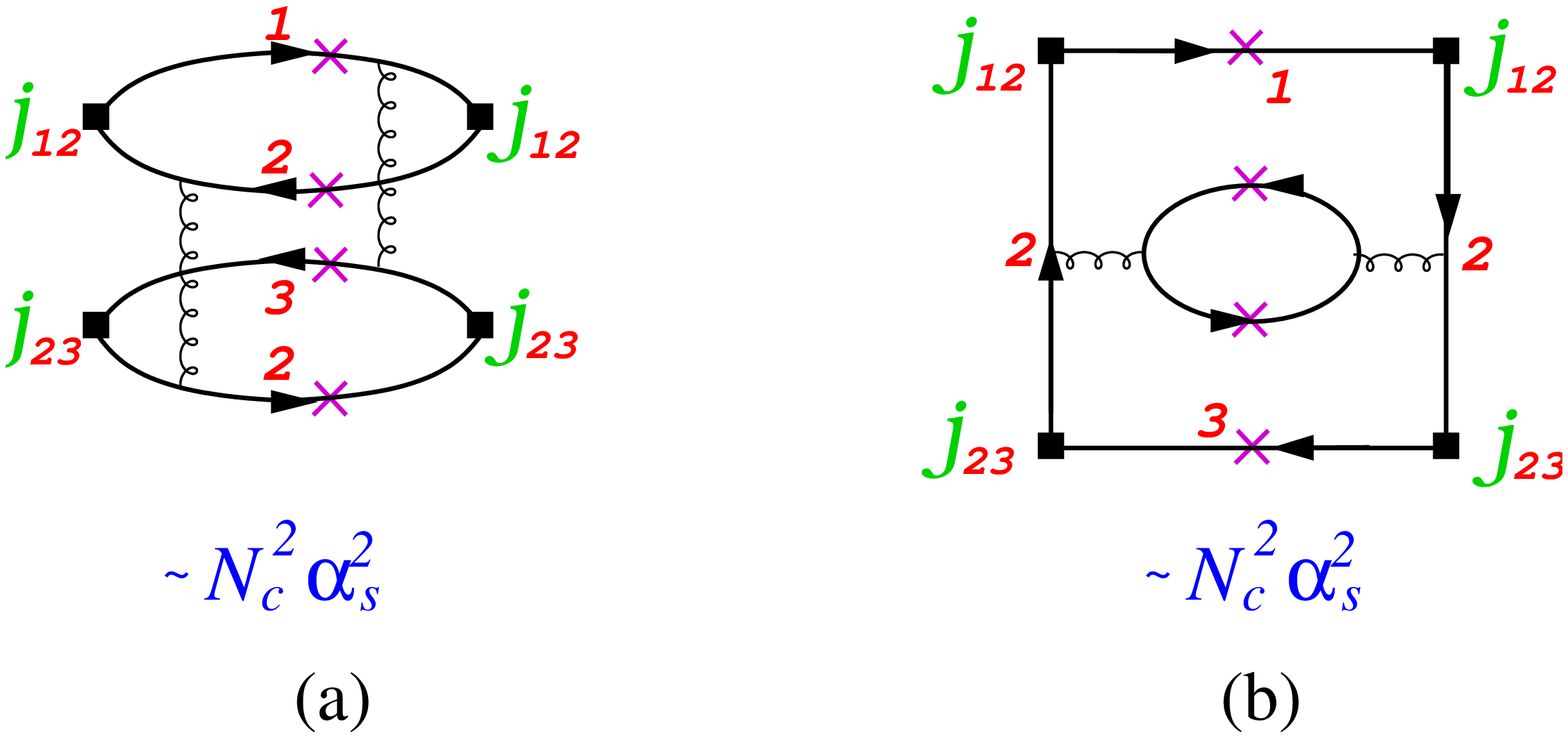}
\caption{\label{Fig:crypto_dir}Diagrams contributing to
$\Gamma_{{\rm I},T}^{\rm(dir)}$. Crossed propagators denote
on-shell particles contributing to the four-quark cut.}
\includegraphics[width=8cm]{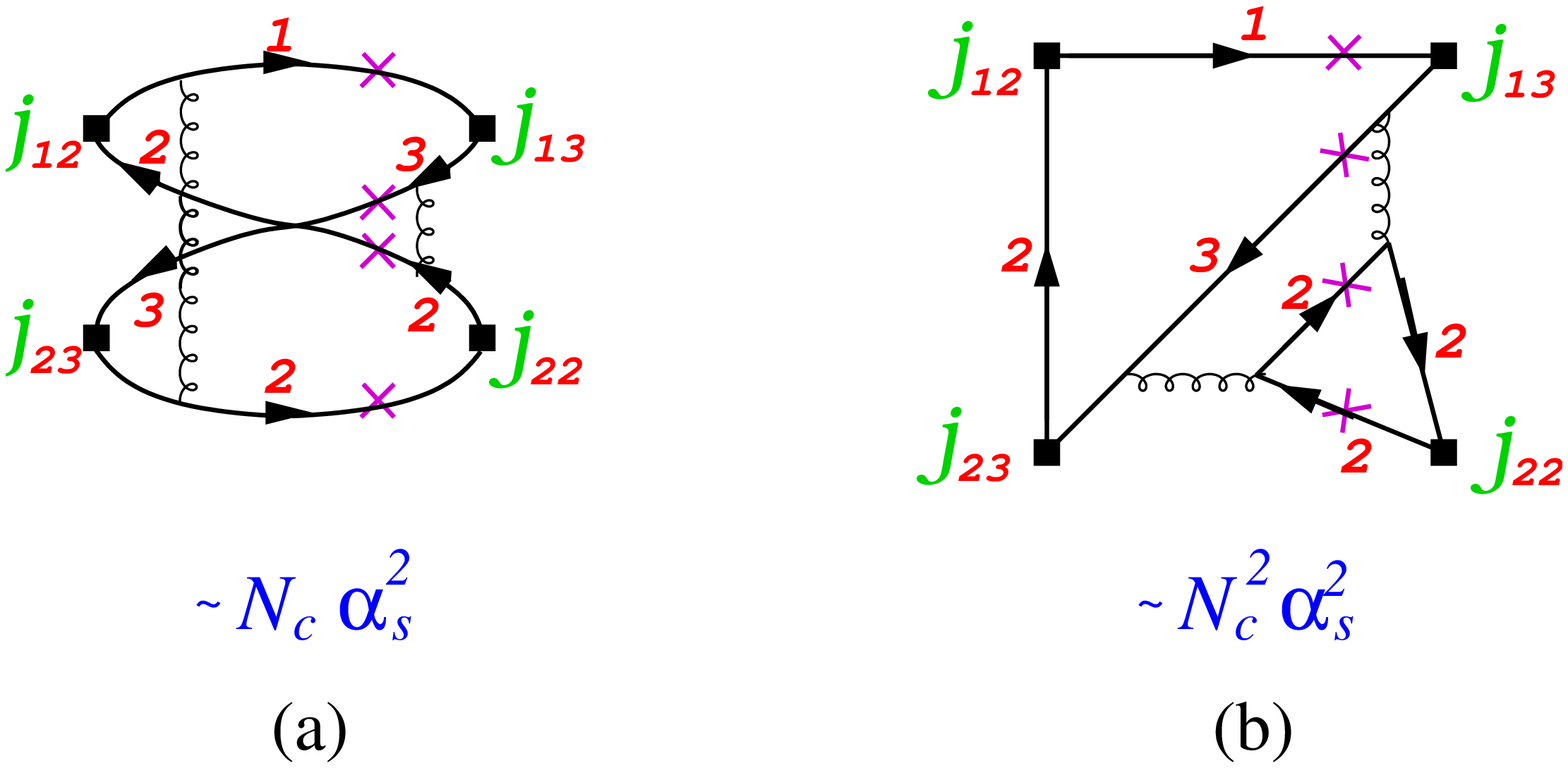}
\caption{\label{Fig:crypto_rec}Diagrams contributing to the
recombination Green function $\Gamma_T^{\rm(rec)}$.}\end{figure}We
now turn to tetraquarks with nonexotic flavor content, i.e.,
having the same flavor as the ordinary mesons. The analysis
proceeds along the same line as for the exotic states. The only
new ingredient is the appearance of~diagrams of new topologies
compared to the flavor-exotic case. For the direct Green functions
$\Gamma^{\rm(dir)}_{({\rm I},{\rm II}),T}$
(Fig.~\ref{Fig:crypto_dir}), the new diagrams do not change the
leading large-$N_c$ behavior compared to the diagrams of the same
topology in the flavor-exotic case.

For the recombination functions, however, the situation changes
qualitatively: the new diagram, Fig.~\ref{Fig:crypto_rec}(b),
dominates at large $N_c$ and thus modifies the leading large-$N_c$
behavior of $\Gamma^{\rm(rec)}_{T}$. We thus find
\begin{eqnarray}\label{C1}
\Gamma^{\rm(dir)}_{{\rm I},T}=\langle
j^\dagger_{12}j^\dagger_{23}j_{12}j_{23}\rangle=O(N_c^0),\quad
\Gamma^{\rm(dir)}_{{\rm II},T}=\langle
j^\dagger_{13}j^\dagger_{22}j_{13}j_{22}\rangle=O(N_c^0),\quad
\Gamma^{\rm(rec)}_T=\langle
j^\dagger_{12}j^\dagger_{23}j_{13}j_{22}\rangle=O(N_c^0).
\end{eqnarray}
In contrast to the flavor-exotic case, now both the direct and the
recombination Green functions have the same leading behavior at
large $N_c$. As a consequence, one exotic state $T$ suffices to
satisfy the expected large-$N_c$ behavior of both Green functions.
The couplings of this state to the meson-meson channels are
\begin{eqnarray}\label{C2}
A(T\to M_{12}M_{23})&=&O(N_c^{-1}),\qquad A(T\to
M_{13}M_{22})=O(N_c^{-1}).
\end{eqnarray}
Thus, the width of this single cryptoexotic state is of order
$\Gamma(T)=O(N_c^{-2})$.

If all its quantum numbers allow, $T$ can mix with the ordinary
meson $M_{13}$. The restriction on the mixing parameter
$g_{TM_{13}}$ may be obtained, e.g., from the direct amplitude
\begin{eqnarray}\label{mixing2}
\Gamma_{{\rm I},T}^{(\rm dir)}=O(N_c^0)
=f_M^4\left(\frac{A(M_{12}M_{23}\to T)}{p^2-M^2_{T}}g_{TM_{13}}
\frac{A(M_{13}\to M_{12}M_{23})}{p^2-M^2_{M_{13}}}\right)+\cdots.
\end{eqnarray}
Taking into account that $A(M_{13}\to M_{12}M_{23})\sim
1/\sqrt{N_c}$ \cite{largeNc1,largeNc2}, we obtain $g_{TM_{13}}\le
O(1/\sqrt{N_c})$.

\section{Conclusions}
We formulated a set of rigorous criteria for selecting those
diagrams that are appropriate for the analysis of potential
tetraquark states: In the four-point Green functions, one should
take into account only those contributions which~have four-quark
cuts in the $s$-channel. Using these criteria and requiring that
the narrow poles contribute to the appropriate parts of the Green
functions at leading large-$N_c$ order, we gained the following
insights:\begin{enumerate}\item The large-$N_c$ behavior of
flavor-exotic four-point Green functions (all four quarks of
different flavors $\bar q_1,\bar q_3,q_2, q_4$) is not compatible
with merely one flavor-exotic tetraquark but requires two narrow
states $T_A$ and $T_B$ with~widths $\Gamma(T_A,T_B)=O(N_c^{-2})$.
Each of these tetraquarks dominantly couples to one meson-meson
channel; its coupling to the other meson-meson channel is
suppressed: $A(T_A\to M_{12}M_{34})=O(N_c^{-1})$, $A(T_A\to
M_{14}M_{32})=O(N_c^{-2})$, $A(T_B\to M_{14}M_{32})=O(N_c^{-1})$,
$A(T_B\to M_{12}M_{34})=O(N_c^{-2})$. The parameter describing the
mixing between~the two tetraquarks vanishes for large $N_c$ at
least like $1/N_c$.\item The large-$N_c$ behavior of four-point
Green functions of nonexotic flavor content ($\bar q_1,\bar
q_2,q_2,q_3$) is compatible with the existence of a single narrow
cryptoexotic tetraquark, $T$, with width $\Gamma(T)=O(N_c^{-2})$.
This tetraquark couples parametrically equally to both two-meson
channels: $A(T\to M_{12}M_{23})=O(N_c^{-1})$, $A(T\to
M_{13}M_{22})=O(N_c^{-1})$. If quantum numbers allow, the
cryptoexotic tetraquark $T=\bar q_1q_3\bar q_2 q_2$ mixes with the
ordinary meson $M_{13}=\bar q_1q_3$. The corresponding mixing
parameter vanishes like $1/\sqrt{N_c}$, i.e., slower than that of
the flavor-exotic tetraquarks.\end{enumerate}We would like to
mention that, in principle, there is a possibility that narrow
tetraquarks do exist but appear only in $N_c$-subleading diagrams
with four-quark intermediate states, while they do not contribute
to the $N_c$-leading~topologies. This possibility seems rather
unnatural to us: if such pole exists at all, there should be some
special reason, not evident to us, why it does not appear in the
set of the appropriate $N_c$-leading diagrams. Nevertheless, also
in the $N_c$-subleading topologies one observes a difference in
the large-$N_c$ behavior of the direct and the recombination
diagrams of the flavor-exotic case: for any $N_c$-subleading
topology, the direct diagrams are $N_c$-even, whereas the
recombination~diagrams are $N_c$-odd. Accordingly, if the latter
scenario is realized in nature, one still encounters the necessity
of two flavor-exotic poles, albeit with parametrically smaller
widths at large $N_c$.

\vspace{.5cm}\noindent{\it{\bf Acknowledgements.}} The authors
thank V.~Anisovich, T.~Cohen, M.~Knecht, B.~Moussallam,
O.~Nachtmann, and B.~Stech for valuable discussions.
D.~M.~acknowledges support from the Austrian Science Fund (FWF),
project~P29028.

\end{document}